\documentclass[conference]{IEEEtran}

\usepackage{graphicx}
\usepackage{url}
\usepackage{flushend}

\usepackage{enumitem}

\begin{document}
\title{Feasibility of Fog Computing }

\author{\IEEEauthorblockN{Blesson Varghese, Nan Wang, Dimitrios S. Nikolopoulos}
\IEEEauthorblockA{School of Electronics, Electrical Engg. and Computer Science\\Queen's University Belfast, UK\\
Email: \{varghese, nwang03, d.nikolopoulos\}@qub.ac.uk}
\and
\IEEEauthorblockN{Rajkumar Buyya}
\IEEEauthorblockA{Dept. of Computing and Information Systems\\University of Melbourne, Australia\\
Email: rbuyya@unimelb.edu.au}
}


%


\maketitle

\begin{abstract}
As billions of devices get connected to the Internet, it will not be sustainable to use the cloud as a centralised server. The way forward is to decentralise computations away from the cloud towards the edge of the network closer to the user. This reduces the latency of communication between a user device and the cloud, and is the premise of \textit{`fog computing'} defined in this paper. The aim of this paper is to highlight the feasibility and the benefits in improving the Quality-of-Service and Experience by using fog computing. For an online game use-case, we found that the average response time for a user is improved by 20\% when using the edge of the network in comparison to using a cloud-only model. It was also observed that the volume of traffic between the edge and the cloud server is reduced by over 90\% for the use-case. The preliminary results highlight the potential of fog computing in achieving a sustainable computing model and highlights the benefits of integrating the edge of the network into the computing ecosystem.
\end{abstract}


\IEEEpeerreviewmaketitle

\section{An Overview}
\label{sec:introduction}
The landscape of parallel and distributed computing has significantly evolved over the last sixty years~\cite{paper0a,paper0b,paper0c}. The 1950s saw the advent of mainframes, after which the vector era dawned in the 1970s. The 1990s saw the rise of the distributed computing or massively parallel processing era. More recently, the many-core era has come to light. These have led to different computing paradigms, supporting full blown supercomputers, grid computing, cluster computing, accelerator-based computing and cloud computing. Despite this growth, there continues to be a significant need for more computational capabilities to meet future challenges. 

\begin{figure}[t]
\centering
\includegraphics[width=0.5\textwidth]{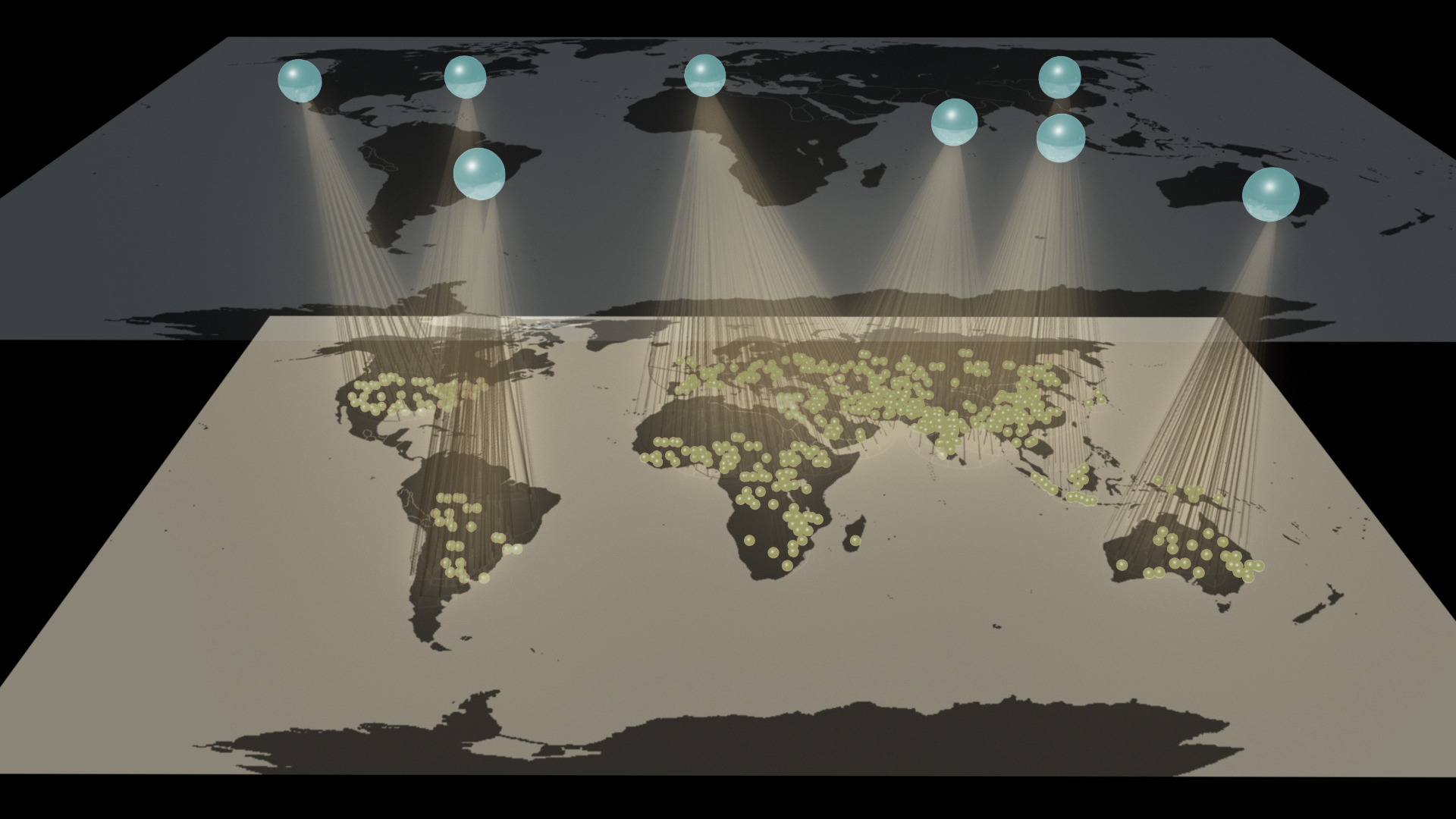}
\caption{A global view of executing applications in the current cloud paradigm where user devices are connected to the cloud. Blue dots show sample locations of cloud data centers and the yellow dots show user devices that make use of the cloud as a centralised server.}
\label{fig:figure1}
\end{figure}

It is forecast that between 20-50 billion devices will be added to the internet by 2020 creating an economy of over \$3~trillion\footnote{\url{http://www.gartner.com/newsroom/id/3165317}}$^{,}$\footnote{\url{http://spectrum.ieee.org/tech-talk/telecom/internet/popular-internet-of-things-forecast-of-50-billion-devices-by-2020-is-outdated}}. Consequently, 43 trillion gigabytes of data will be generated and will need to be processed in cloud data centers. Applications generating data on user devices, such as smartphones, tablets and wearables currently use the cloud as a centralised server (as shown in Figure~\ref{fig:figure1}), but this will soon become an untenable computing model. This is simply because the frequency and latency of communication between user devices and geographically distant data centers will increase beyond that which can be handled by existing communication and computing infrastructure~\cite{paper1}. This will adversely affect Quality-of-Service (QoS) and Quality-of-Experience (QoE)~\cite{paper3}.

\begin{figure}[t]
\centering
\includegraphics[width=0.5\textwidth]{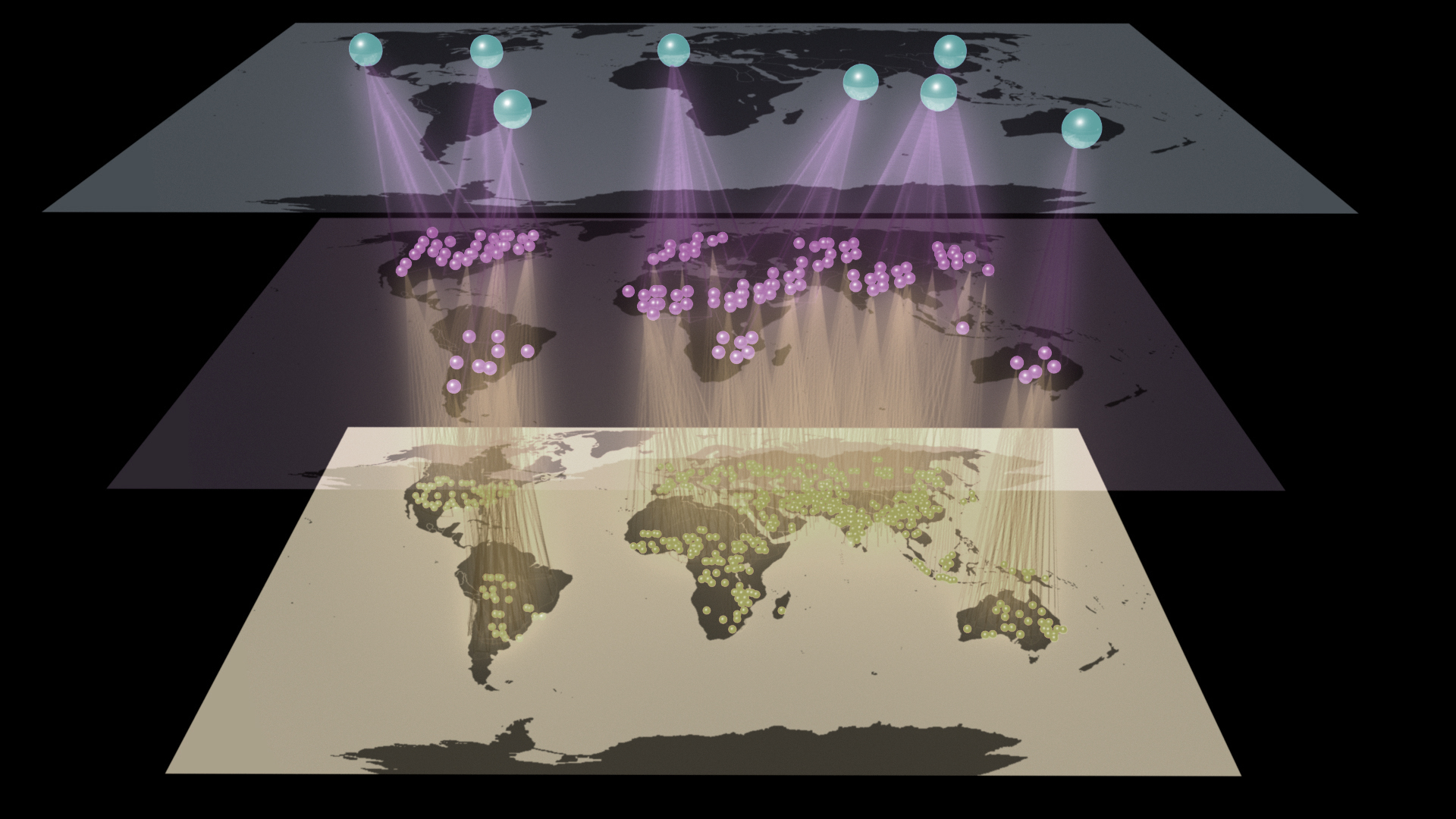}
\caption{A global view of executing applications at the edge of the network in the fog computing model where user devices are connected to the cloud indirectly. The user devices are serviced by the edge nodes. Blue dots show sample locations of cloud data centers and the yellow dots show user devices that make use of the cloud through a variety of edge nodes indicated in purple.}
\label{fig:figure2}
\end{figure}

Applications will need to process data closer to its source to reduce network traffic and efficiently deal with the data explosion. However, this may not be possible on user devices, since they have relatively restricted hardware resources. Hence, there is strong motivation to look beyond the cloud towards the edge of the network to harness computational capabilities that are currently untapped~\cite{agarwal2014vision,meurisch2015upgrading}. For example, consider routers, mobile base stations and switches that route network traffic. The computational resources available on such nodes, referred to as \textit{`Edge Nodes'} that are situated closer to the user device than the data center can be employed. 

We define the concept of distributed computing on the edge of the network in conjunction with the cloud, referred to as \textit{`Fog Computing'}~\cite{fogcomputing-0,fogcomputing-1,fogcomputing-2}. This computing model is based on the premise that computational workloads can be executed on edge nodes situated in between the cloud and a host of user devices to reduce communication latencies and offer better QoS and QoE as shown in Figure~\ref{fig:figure2}.
In this paper, we refer to edge nodes as the nodes located at the edge of the network whose computational capabilities are harnessed. This model co-exists with cloud computing to complement the benefits offered by the cloud, but at the same time makes computing more feasible as the number of devices increases. 

We differentiate this from `edge computing'~\cite{paper1,paper3,paper2} in which the edge of the network, for example, nodes that are one hop away from a user device, is employed only for complementing computing requirements of user devices. On the other hand, in fog computing, computational capabilities across the entire path taken by data may be harnessed, including the edge of the network. Both computing models use the edge node; the former integrates it in the computing model both with the cloud and user devices, where as the latter incorporates it only for user devices. 

In this paper, we provide a definition of fog computing and articulate its distinguishing characteristics. Further, we provide a view of the computing ecosystem that takes the computing nodes, execution models, workload deployment techniques and the marketplace into account. A location-aware online game use-case is presented to highlight the feasibility of fog computing. The average response time for a user is improved by 20\% when compared to a cloud-only model. Further, we observed a 90\% reduction in data traffic between the edge of the network and the cloud. The key result is that the fog computing model is validated.

The remainder of this paper is organised as follows.
Section~\ref{sec:definitions} define fog computing and presents characteristics that are considered in the fog computing model.
Section~\ref{sec:architecture} presents the computing ecosystem, including the nodes, workload execution, workload deployment, the fog marketplace.
Section~\ref{sec:experiments} highlights experimental results obtained from comparing the cloud computing and fog computing models. 
Section~\ref{sec:conclusions} concludes this paper. 

\section{Definition and Characteristics of\\Fog Computing}
\label{sec:definitions}
A commonly accepted definition for cloud computing was provided by the National Institute for Standards and Technology (NIST) in 2011, which  was ``... a model for enabling ubiquitous, convenient, on-demand network access to a shared pool of configurable computing resources (e.g., networks, servers, storage, applications, and services) that can be rapidly provisioned and released with minimal management effort or service provider interaction.'' This definition is complemented by definitions provided by IBM\footnote{\url{https://www.ibm.com/cloud-computing/what-is-cloud-computing}} and Gartner\footnote{\url{http://www.gartner.com/it-glossary/cloud-computing/}}. The key concepts that are in view are on-demand services for users, rapid elasticity of resources and measurable services for transparency and billing~\cite{intro-1,intro-2,academiccloudcomp-1}.

\subsection{Definition}
We define \textit{\textbf{fog computing} as a model to complement the cloud for decentralising the concentration of computing resources (for example, servers, storage, applications and services) in data centers towards users for improving the quality of service and their experience}. 

In the fog computing model, computing resources already available on weak user devices or on nodes that are currently not used for general purpose computing may be used. Alternatively, additional computational resources may be added onto nodes one or a few hops away in the network to facilitate computing closer to the user device. This impacts latency, performance and quality of the service positively~\cite{latency-1,distribute-1}. This model in no way can replace the benefits of using the cloud, but optimises performance of applications that are user-driven and communication intensive. 

Consider for example, a location-aware online game use-case that will be presented in Section~\ref{sec:experiments}. Typically, such a game would be hosted on a cloud server and the players connect to the server through devices, such as smartphones and tablets. Since the game is location-aware the GPS coordinates will need to be constantly updated based on the players movement. This is communication intensive. The QoS may be affected given that the latency between a user device and a distant cloud server will be high. However, if the game server can be brought closer to the user, then latency and communication frequency can be reduced. This will improve the QoS and QoE.    
The fog computing model can also incorporate a wide variety of sensors to the network without the requirement of communicating with distant resources, thereby allowing low latency actuation efficiently~\cite{sensor-1,sensor-2}. For example, sensor networks in smart cities generating large volumes of data can be processed closer to the source without transferring large amounts of data across the internet.   

Another computing model that is sometimes synonymously used in literature is edge computing~\cite{paper1,paper3,paper2}. We distinguish fog computing and edge computing in this paper. In edge computing, the edge of the network (for example, nodes that are one hop away from a user device) is employed for only facilitating computing of user devices. In contrast, the aim in fog computing is to harness computing across the entire path taken by data, which may include the edge of the network closer to a user. Computational needs of user devices and edge nodes can be complemented by cloud-like resources that may be closer to the user or alternatively workloads can be offloaded from cloud servers to the edge of the network. Both the edge and fog computing models complement each other and given the infancy of both computing models, the distinctions are not obvious in literature.  

\subsection{Characteristics}
Cloud concepts, such as on-demand services for users, rapid elasticity of resources and measurable services for transparency will need to be achieved in fog computing. The following characteristics specific to fog computing will need to be considered in addition:

\subsubsection{\textbf{Vertical Scaling}}
Cloud data centers enable on-demand resource scaling horizontally. Multiple Virtual Machines (VMs), for example, could be employed to meet the increasing requests made by user devices to a web server during peak hours (horizontal scaling is facilitated by the cloud). However, the ecosystem of fog computing will offer resource scaling vertically, whereby multiple hierarchical levels of computations offered by different edge nodes could be introduced to not only reduce the amount of traffic from user devices that reaches the cloud, but also reduce the latency of communication. Vertical scaling is more challenging since resources may not be tightly coupled as servers in a data center and may not necessarily be under the same ownership. 

\subsubsection{\textbf{Heterogeneity}}
On the cloud, virtual resources are usually made available across homogeneous physical machines. For example, a specific VM type provided by Amazon is mapped on to the same physical server. On the other hand, the fog computing ecosystem comprises heterogeneous nodes ranging from sensors to user devices to routers, mobile base stations and switches to large machines situated in data centers. These devices and nodes have CPUs with varying specifications and performance capabilities, including Digital Signal Processors (DSPs) or other accelerators, such as Graphics Processing Units (GPUs). Facilitating general purpose computing on such a variety of resources both at the horizontal and vertical scale is the vision of fog computing.  

\subsubsection{\textbf{Visibility and Accessibility}}
Resources in the cloud are made publicly accessible and are hence visible to a remote user through a marketplace. The cloud marketplace is competitive and makes a wide range of offerings to users. 
In fog computing, a significantly larger number of nodes in the network that would not be otherwise visible to a user will need to become publicly accessible. Developing a marketplace given the heterogeneity of resources and different ownership will be challenging. Moreover, building consumer confidence in using fog enabled devices and nodes will require addressing a number of challenges, such as security and privacy, developing standards and benchmarks and articulating risks. 

\subsubsection{\textbf{Volume}}
There is an increasing number of resources that are added to a cloud data center to offer services. With vertical scaling and heterogeneity as in fog computing the number of resources that will be added and that will become visible in the network will be large. As previously indicated, billions of devices are expected to be included in the network. In addition to a vertical scale out, a horizontal scale out is inevitable.

\section{The Fog Computing Ecosystem}
\label{sec:architecture}
\begin{figure}[t]
\centering
\includegraphics[width=0.49\textwidth]{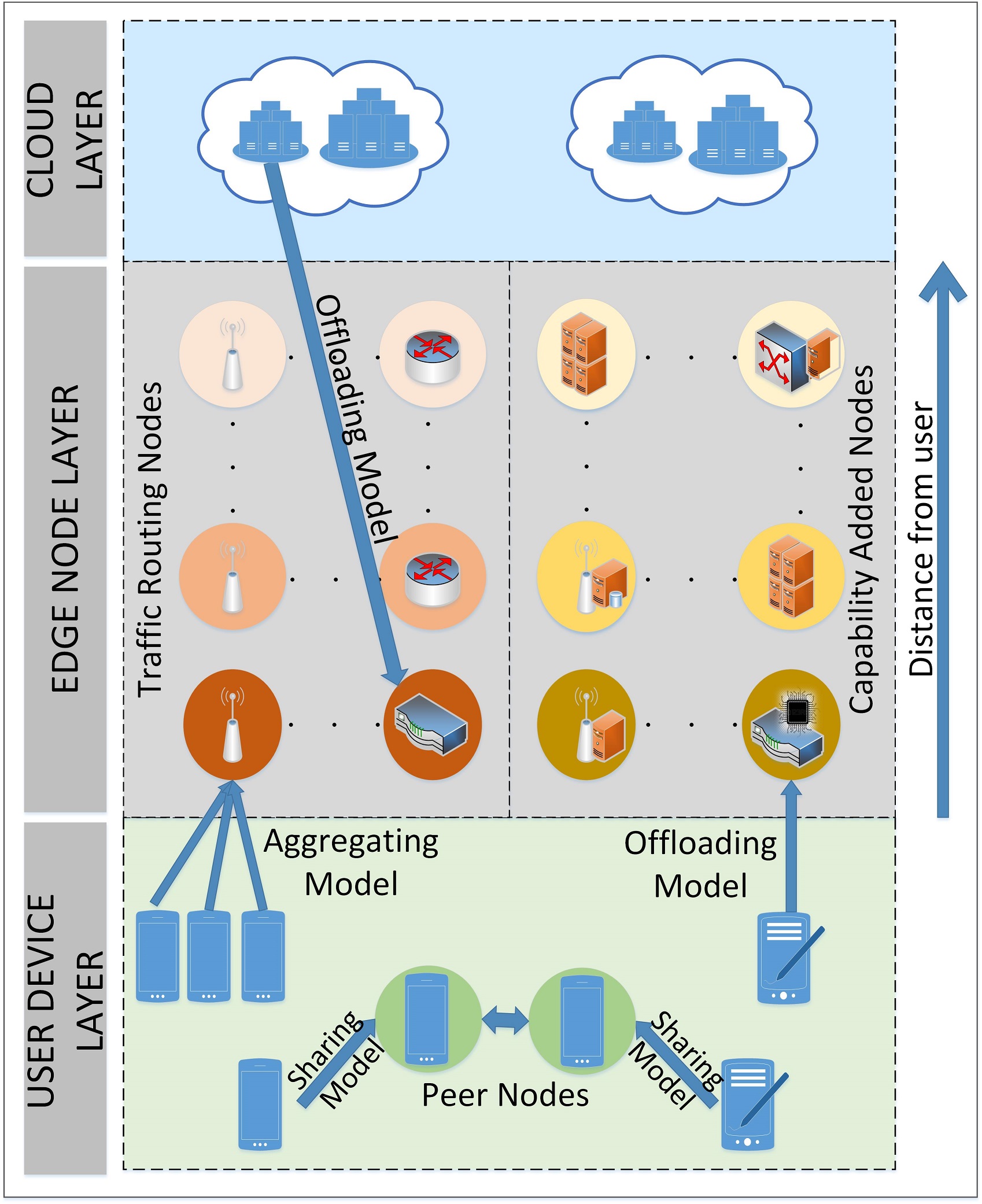}
\caption{The fog computing ecosystem considered in Section~\ref{sec:architecture} showing the user device layer, edge node layer and cloud layer. The user device layer comprises user devices that would traditionally communicate with the cloud. The edge node layer comprises multiple hierarchical levels of edge nodes. However, in the fog computing model, nodes close to the user are of particular interest since the aim is to bring computing near user devices where data is generated. The different nodes, include traffic routing nodes (such as base stations, routers, switches and gateways), capability added nodes (such as traffic routing nodes, with additional computational capabilities, or dedicated computational resources) and peer nodes (such as a collection of volunteered user devices as a dynamic cloud). Workloads are executed in an offloading (both from user device to the edge and from the cloud to the edge), aggregating and sharing models (or a hybrid combining the above, which is not shown) on edge nodes closer to the user.}
\label{fig:figureedgecomputingarchitecture}
\end{figure}

In the cloud-only computing model, the user devices at the edge of the network, such as smartphones, tablets and wearables, communicate with cloud servers via the internet as shown in Figure~\ref{fig:figure1}. Data from the devices are stored in the cloud. All communication is facilitated through the cloud, as if the devices were talking to a centralised server. Computing and storage resources are concentrated in the cloud data centers and user devices simply access these services. For example, consider a web application that is hosted on a server in a data center or multiple data centers. Users from around the world access the web application service using the internet. The cloud resource usage costs are borne by the company offering the web application and they are likely to generate revenue from users through subscription fees or by advertising.

However in the fog computing model as shown in Figure~\ref{fig:figureedgecomputingarchitecture}, computing is not only concentrated in cloud data centers. Computation and even storage is brought closer to the user, thus reducing latencies due to communication overheads with remote cloud servers~\cite{offloading-1,offloading-2,offloading-4}. This model aims to achieve geographically distributed computing by integrating multiple heterogeneous nodes at the edge of the network that would traditionally not be employed for computing.

\subsection{Computing Nodes}
Typically, CPU-based servers are integrated to host VMs in the cloud.  
Public clouds, such as the Amazon Elastic Compute Cloud (EC2)\footnote{\url{https://aws.amazon.com/ec2/}} or the Google Compute Engine\footnote{\url{https://cloud.google.com/compute/}} offer VMs through dedicated servers that are located in data centers. Hence, multiple users can share the same the physical machine. Private clouds, such as those owned by individual organisations, offer similar infrastructure but are likely to only execute workloads of users from within the organisation. 

To deliver the fog computing vision, the following nodes will need to be integrated in the computing ecosystem:
\subsubsection{\textbf{\textit{Traffic routing nodes}}}
through which the traffic of user devices is routed (those that would not have been traditionally employed for general purpose computing), such as routers, base stations and switches. 
\subsubsection{\textbf{\textit{Capability added nodes}}} by extending the existing capabilities of traffic routing nodes with additional computational and storage hardware or by using dedicated compute nodes.
\subsubsection{\textbf{\textit{Peer nodes}}} which may be user devices or nodes that have spare computational cycles and are made available in the network as volunteers or in a marketplace on-demand.

Current research aims to deliver fog computing using private clouds. The obvious advantage in limiting visibility of edge nodes and using proprietary architectures is bypassing the development of a public marketplace and its consequences. 
Our vision is that in the future, the fog computing ecosystem will incorporate both public and private clouds. This requires significant research and development to deliver a marketplace that makes edge nodes publicly visible similar to public cloud VMs. Additionally, technological challenges in managing resources and enhancing security will need to be accounted for. 

\subsection{Workload Execution Models}
Given a workload, the following execution models can be adopted on the fog ecosystem for maximising performance.

\subsubsection{\textbf{\textit{Offloading model}}}
Workloads can be offloaded in the following two ways. Firstly, from user devices onto edge nodes to complement the computing capabilities of the device. For example, consider a face or object recognition application that may be running on a user device. This application may execute a parallel algorithm and may require a large number of computing cores to provide a quick response to the user. In such cases, the application may offload the workload from the device onto an edge node, for example a capability added node that comprises hardware accelerators or many cores. 

Secondly, from cloud servers onto edge nodes so that computations can be performed closer to the users. Consider for example, a location aware online game to which users are connected from different geographic locations. If the game server is hosted in an Amazon data center, for example in N. Virginia, USA, then the response time for European players may be poor. The component of the game server that services players can be offloaded onto edge nodes located closer to the players to improve QoS and QoE for European players. 

\subsubsection{\textbf{\textit{Aggregating model}}}
Data streams from multiple devices in a given geographic area are routed through an edge that performs computation, to either respond to the users or route the processed data to the cloud server for further processing. For example, consider a large network of sensors that track the level of air pollution in a smart city. The sensors may generate large volumes of data that do not need to be shifted to the cloud. Instead, edge nodes may aggregate the data from different sensors, either to filter or pre-process data, and then forward them further on to a more distant server. 

\subsubsection{\textbf{\textit{Sharing model}}}
Workloads relevant to user devices or edge nodes are shared between peers in the same or different hierarchical levels of the computing ecosystem. For example, consider a patient tracking use-case in a hospital ward. The patients may be supplied wearables or alternate trackers that communicate with a pilot device, such as a chief nurse's smartphone used at work. Alternatively, the data from the trackers could be streamed in an aggregating model. Another example includes using compute intensive applications in a bus or train. Devices that have volunteered to share their resources could share the workload of a compute intensive application.

\subsubsection{\textbf{\textit{Hybrid model}}}
Different components of complex workloads may be executed using a combination of the above strategies to optimise execution. Consider for example, air pollution sensors in a city, which may have computing cores on them. When the level of pollutants in a specific area of the city is rising, the monitoring frequency may increase resulting in larger volumes of data. This data could be filtered or pre-processed on peer nodes in the sharing model to keep up with the intensity at which data is generated by sensors in the pollution high areas. However, the overall sensor network may still follow the aggregating model considered above.  

\subsection{Workload Deployment Technologies}
While conventional Operating Systems (OS) will work on large CPU nodes, micro OS that are lightweight and portable may be suitable on edge nodes. Similar to the cloud, abstraction is key to deployment of workloads on edge nodes~\cite{virtualisation-1}. Technologies that can provide abstraction are:

\subsubsection{\textbf{\textit{Containers}}} 
The need for lightweight abstraction that offers reduced boot up times and isolation is offered by containers. Examples of containers include, Linux containers~\cite{linuxcontainers-1} at the OS level and Docker~\cite{docker-1} at the application level. 

\subsubsection{\textbf{\textit{Virtual Machines (VMs)}}} 
On larger and dedicated edge nodes that may have substantial computational resources, VMs provided in cloud data centers can be employed. 

These technologies have been employed on cloud platforms and work best with homogeneous resources. The heterogeneity aspect of fog computing will need to be considered to accommodate a wider range of edge nodes. 

\subsection{The Marketplace}
The public cloud marketplace has become highly competitive and offers computing as a utility by taking a variety of CPU, storage and communication metrics into account~\cite{cloudpricing-1,cloudpricing-2}. For example, Amazon's pricing of a VM is based on the number of virtual CPUs and memory allocated to the VM. To realise fog computing as a utility, a similar yet a more complex marketplace will need to be developed. The economics of this marketplace will be based on:

\subsubsection{\textbf{\textit{Ownership}}} 
Typically, public cloud data centers are owned by large businesses. If traffic routing nodes were to be used as edge nodes, then their ownership is likely to be telecommunication companies or governmental organisations that may have a global reach or are regional players (specific to the geographic location. For example, a local telecom operator). Distributed ownership will make it more challenging to obtain a unified marketplace operating on the same standards. 

\subsubsection{\textbf{\textit{Pricing Models}}} 
On the edge there are three possible levels of communication, which are between the user devices and the edge node, one edge node and another edge node, and an edge node and a cloud server, which will need to be considered in a pricing model. In addition, `who pays what' towards the bill has to be articulated and a sustainable and transparent economic model will need to be derived. Moreover, the priority of applications executing on these nodes will have to be accounted for. 

\subsubsection{\textbf{\textit{Customers}}} 
Given that there are multiple levels of communication when using an edge node, there are potentially two customers. The first is an application owner running the service on the cloud who wants to improve the quality of service for the application user. For example, in the online game use-case considered previously, the company owning the game can improve the QoS for customers in specific locations (such as Oxford Circus in London that and Times Square in New York that is often crowded) by hosting the game server on multiple edge node locations. This will significantly reduce the application latency and may satisfy a large customer base.

The second is the application user who could make use of an edge node to improve the QoE of a cloud service via fog computing. Consider for example, the basic services of an application on the cloud that are currently offered for free. A user may choose to access the fog computing based service of the application for a subscription or on a pay-as-you-go basis to improve the user experience, which is achieved by improving the application latency. 

For both the above, in addition to existing service agreements, there will be requirements to create agreements between the application owner, the edge node and the user, which can be transparently monitored within the marketplace. 

\subsection{Other concepts to consider}
While there are a number of similarities with the cloud, fog computing will open a number of avenues that will make it different from the cloud. The following four concepts at the heart of fog computing will need to be approached differently than current implementations on the cloud:

\subsubsection{\textbf{Priority-based multi-tenancy}}
In the cloud, multiple VMs owned by different users are co-located on the same physical server~\cite{cloudmultitenancy-1,cloudmultitenancy-2}. These servers unlike many edge nodes are reserved for general purpose computing. Edge nodes, such as a mobile base station, for example, are used for receiving and transmitting mobile signals. The computing cores available on such nodes are designed and developed for the primary task of routing traffic. However, if these nodes are used in fog computing and if there is a risk of compromising the QoS of the primary service, then a priority needs to be assigned to the primary service when co-located with additional computational tasks. Such priorities are usually not required on dedicated cloud servers. 

\subsubsection{\textbf{Complex Management}}
Managing a cloud computing environment requires the fulfilment of agreements between the provider and the user in the form of Service Level Agreements (SLAs)~\cite{sla-1,sla-2}. This becomes complex in a multi-cloud environment~\cite{broker-2,broker-4}. However, management in fog computing will be more complex given that edge nodes will need to be accessible through a marketplace. If a task were to be offloaded from a cloud server onto an edge node, for example, a mobile base station owned by a telecommunications company, then the cloud SLAs will need to take into account agreements with a third-party. Moreover, the implications to the user will need to be articulated. The legalities of SLAs binding both the provider and the user in cloud computing are continuing to be articulated. Nevertheless, the inclusion of a third party offering services and the risk of computing on a third party node will need to be articulated. Moreover, if computations span across multiple edge nodes, then monitoring becomes a more challenging task. 

\subsubsection{\textbf{Enhanced Security and Privacy}}
The key to computing remotely is security that needs to be guaranteed by a provider~\cite{cloudsecurity-1,cloudsecurity-2}. In the cloud context, there is significant security risk related to data storage and hosting multiple users. Robust mechanisms are currently offered on the cloud to guarantee user and user data isolation. This becomes more complex in the fog computing ecosystem, given that not only are the above risks of concern, but also the security concerns around the traffic routed through nodes, such as routers~\cite{edgesecurity-1,edgesecurity-2}. For example, a hacker could deploy malicious applications on an edge node, which in turn may exploit a vulnerability that may degrade the QoS of the router. Such threats may have a significant negative impact. Moreover, if user specific data needs to be temporarily stored on multiple edge locations to facilitate computing on the edge, then privacy issues along with security challenges will need to be addressed. Vulnerability studies that can affect security and privacy of a user on both the vertical and horizontal scale will need to be freshly considered in light of facilitating computing on traffic routing nodes. 

\subsubsection{\textbf{Lighter Benchmarking and Monitoring}}
Performance is measured on the cloud using a variety of techniques, such as benchmarking to facilitate the selection of resources that maximise performance of an application and periodic monitoring of the resources to ensure whether user-defined service level objectives are achieved~\cite{benchmarking-2,benchmarking-4,benchmarking-5}. Existing techniques are suitable in the cloud context since they monitor nodes that are solely used for executing the workloads~\cite{monitoring-0,monitoring-1,monitoring-2}. On edge nodes however, monitoring will be more challenging, given the limited hardware availability. Secondly, benchmarking and monitoring will need to take into account the primary service, such as routing traffic, that cannot be compromised. Thirdly, communication between the edge node and user devices and the edge node and the cloud and potential communication between different edge nodes will need to be considered. Fourthly, vertical scaling along multiple hierarchical levels and heterogeneous devices will need to be considered. These are not important considerations on the cloud, but become significant in the context of fog computing.  

\section{Preliminary Results}
\label{sec:experiments}
In this section, we present preliminary results that indicate that fog computing is feasible and in using the edge of the network in conjunction with the cloud has potential benefits that can improve QoS and QoE. The use-case employed is an open-sourced version of a location-aware online game similar to Pok\'eMon Go, named iPokeMon\footnote{\url{https://github.com/Kjuly/iPokeMon}}. The game features a virtual reality environment that can be played on a variety of devices, such as smartphones and tablets. The user locates, captures, battles and trains virtual reality creatures, named Pok\'emons, through the GPS capability of the device. 
The Pok\'emons are geographically distributed and a user aims to build a high value profile among their peers. The users may choose to walk or jog through a city to collect Pok\'emons.    

The current execution model, which is a \textbf{\textit{cloud-only model}}, is such that the game server is hosted on the public cloud and the users connect to the server. The server updates the user position and a global view of each user and the Pok\'emons is maintained by the server. For example, if the Amazon EC2 servers are employed, then the game may be hosted in the EC2 N. Virginia data center and a user in Belfast (over 3,500 miles) communicates with the game server. This may be optimised by the application owner in hosting the server closer to Belfast in the Dublin data center (which is nearly a 100 miles away from the user). The original game server is known to have crashed multiple times during its launch due to severe activities which were not catered for\footnote{\url{http://www.forbes.com/sites/davidthier/2016/07/07/pokemon-go-servers-seem-to-be-struggling/#588a88b64958}}$^{,}$\footnote{\url{https://www.theguardian.com/technology/2016/jul/12/pokemon-go-australian-users-report-server-problems-due-to-high-demand}}.

\begin{figure}
\centering
\includegraphics[width=0.495\textwidth]{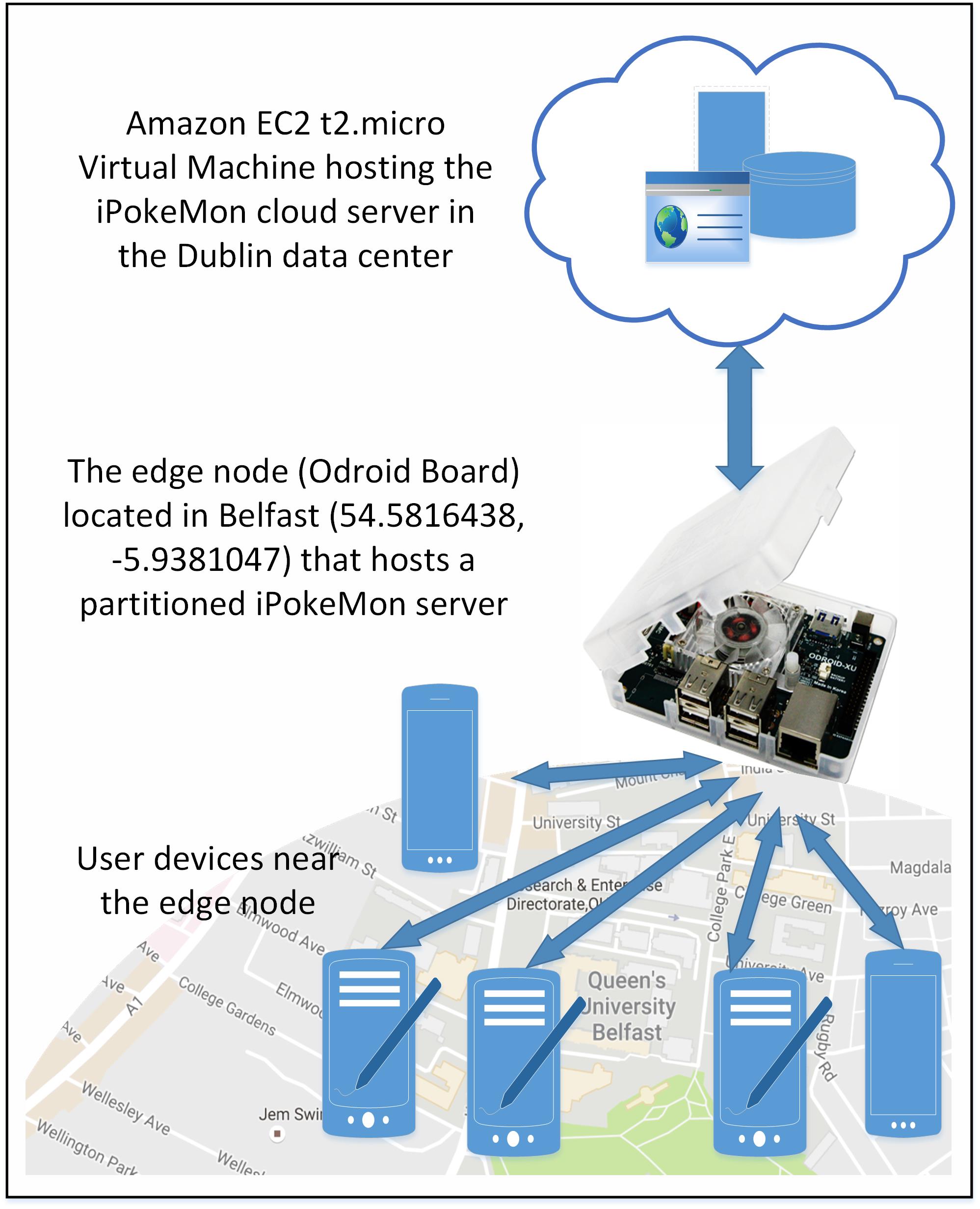}
\caption{The experimental testbed used for implementing the fog computing-based iPokeMon game. The cloud server was hosted in the Amazon Dublin data center on a t2.micro virtual machine. The server on the edge of the network was hosted on the Odroid board, which was located in Belfast. Multiple game clients that were in close proximity to the edge node established connection with the edge server to play the game.}
\label{fig:gameonedge}
\end{figure}

We implemented an \textbf{\textit{fog computing model}} for executing the iPokeMon game. The data packets sent from a smartphone to the game server will pass through a \textit{traffic routing node}, such as a mobile base station. We assumed a mobile base station (the edge node) was in proximity of less than a kilometre to a set of iPokeMon users. Modern base stations have on-chip CPUs, for example the Cavium Octeon Fusion processors\footnote{\url{http://www.cavium.com/OCTEON-Fusion.html}}. Such processors have between 2 and 6 CPU cores with between 1 to 2 GB RAM memory to support between 200-300 users. To represent such a base station we used an ODROID-XU+E board\footnote{\url{http://www.hardkernel.com/}}, which has similar computing resources as a modern base station. The board has one ARM Big.LITTLE architecture Exynos~5~Octa processor and 2~GB of DRAM memory. The processor runs Ubuntu~14.04 LTS. 

We partitioned the game server to be hosted on both the Amazon EC2 Dublin data center\footnote{\url{https://aws.amazon.com/about-aws/global-infrastructure/}} in the Republic of Ireland and our edge node located in the Computer Science Building of the Queen's University Belfast\footnote{\url{http://www.qub.ac.uk}} in Northern Ireland. The cloud server was hosted on a t2.micro instance offered by Amazon and the server on the edge node was hosted using Linux containers. Partitioning was performed, such that the cloud server maintained a global view of the Pok\'emons, where as the edge node server had a local view of the users that were connected to the edge server. The edge node periodically updated the global view of the cloud server. Resource management tasks in fog computing involving provisioning of edge nodes and auto-scaling of resources allocated to be taken into account. The details of the fog computing-based implementation are beyond the scope of this paper presenting preliminary results and will be reported elsewhere.

\begin{figure}
\centering
\includegraphics[width=0.5\textwidth]{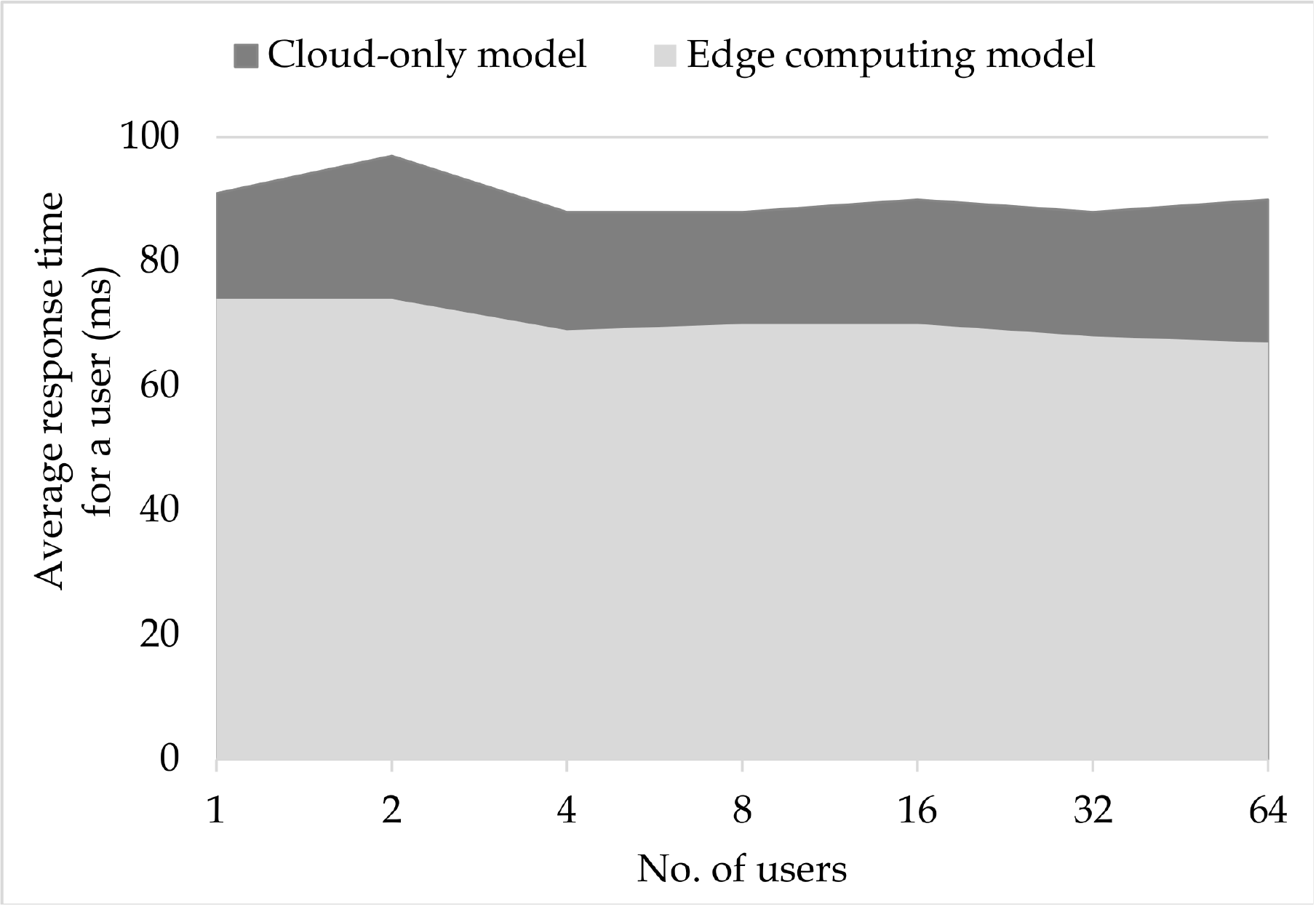}
\caption{Comparing average response time of iPokeMon game users when using a server located on the cloud and on an edge node. In the fog computing model, an improvement of over 20\% is noted when the server is located on the edge node.}
\label{fig:figure3}
\end{figure}

Figure~\ref{fig:figure3} shows the average response time from the perspective of a user, which is measured by round trip latency from when the user device generates a request while playing the game that needs to be serviced by a cloud server (this includes the computation time on the server). The response time is noted over a five minute time period for varying number of users. In the fog computing model, it is noted that on an average the response time can be reduced in the edge computing model for the user playing the game by over 20\%.

\begin{figure}
\centering
\includegraphics[width=0.5\textwidth]{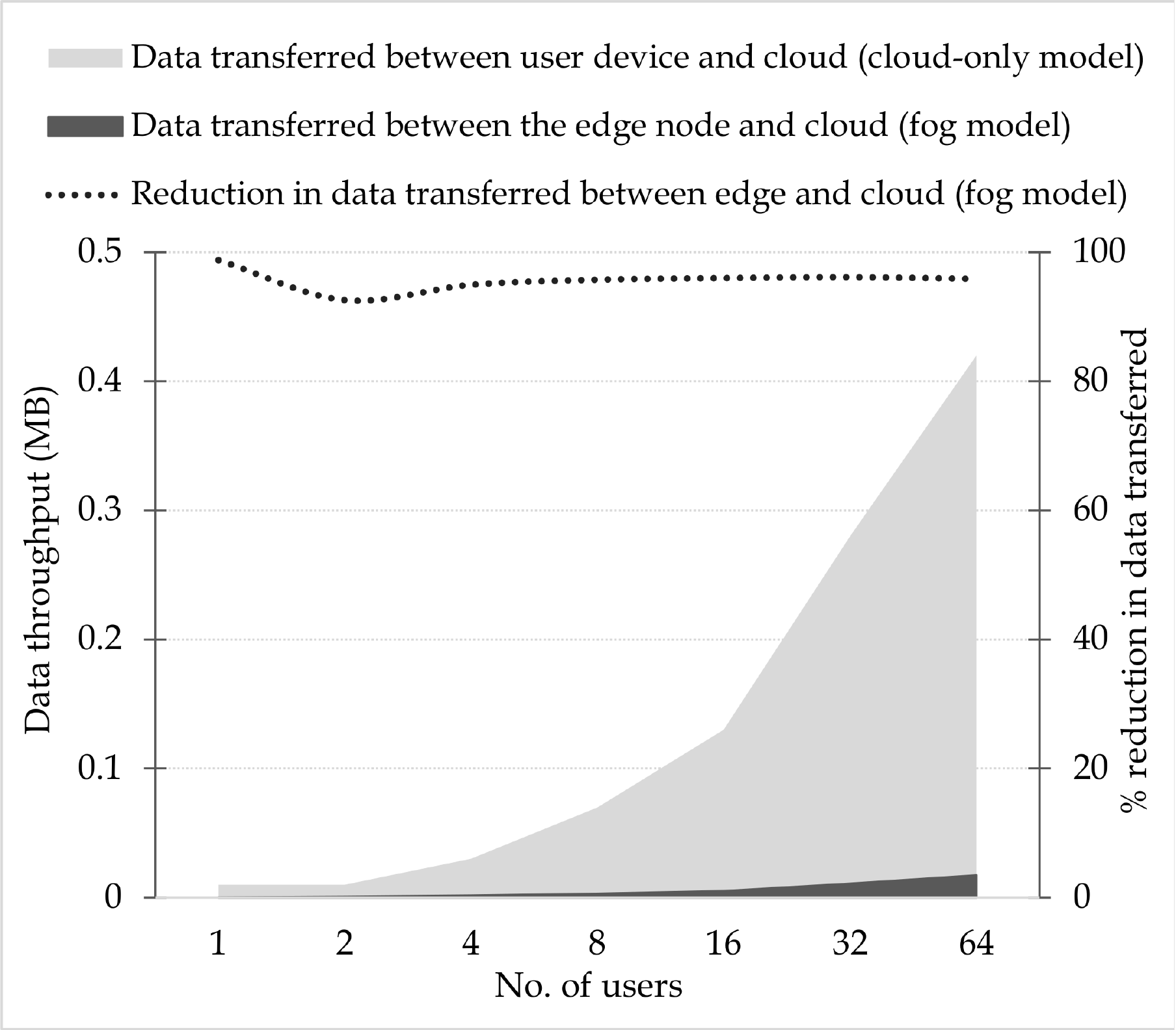}
\caption{Percentage reduction in the data traffic between edge nodes and the cloud to highlight the benefit of using the fog computing model. The data transferred between the edge node and the cloud is reduced by 90\%.}
\label{fig:figure4}
\end{figure}

Figure~\ref{fig:figure4} presents the amount of data that is transferred during the five minute time period to measure the average response time. As expected with increasing number of users the data transferred increases. However, we observe that in the fog computing model the data transferred between the edge node and the cloud is significantly reduced, yielding an average of over 90\% reduction. 

The preliminary results for the given online game use-case highlight the potential of using fog computing in reducing the communication frequency between a user device and a remote cloud server, thereby improving the QoS and QoE.

\section{Conclusions}
\label{sec:conclusions}
The fog computing model can reduce the latency and frequency of communication between a user and an edge node. This model is possible when concentrated computing resources located in the cloud are decentralised towards the edge of the network to process workloads closer to user devices. In this paper, we have defined fog computing and contrasted it with the cloud. An online game use-case was employed to test the feasibility of the fog computing model. The key result is that the latency of communication decreases for a user thereby improving the QoS when compared to a cloud-only model. Moreover, it is observed that the amount of data that is transferred towards the cloud is reduced. 

Fog computing can improve the overall efficiency and performance of applications. These benefits are currently demonstrated on research use-cases and there are no commercial fog computing services that integrate the edge and the cloud models. There are a number of challenges that will need to be addressed before this integration can be achieved and fog computing can be delivered as a utility~\cite{edgecomputing-00}. First of all, a marketplace will need to be developed that makes edge nodes visible and accessible in the fog computing model. This is not an easy task, given that the security and privacy concerns in using edge nodes will need to be addressed. Moreover, potential edge node owners and cloud service providers will need to come to agreement on how edge nodes can be transparently monitored and billed in the fog computing model. To this end, standards and benchmarks will need to be developed, pricing models will need to take multiple party service level agreements and objectives into account, and the risk for the user will need to be articulated. Not only are these socio-economic factors going to play an important role in the integration of the edge and the cloud in fog computing, but from the technology perspective, workload deployment models and associated programming languages and tool-kits will need to be developed.


\bibliographystyle{IEEEtran}  
\bibliography{references}

\end{document}